# RESPONSE OF AN IMPACTING HERTZIAN CONTACT TO AN ORDER-2 SUBHARMONIC EXCITATION : THEORY AND EXPERIMENTS.

J. Perret-Liaudet*, E. Rigaud

*École Centrale de Lyon*

*Laboratoire de Tribologie et Dynamique des Systèmes, UMR 5513,*

*36 Avenue Guy de Collongue, F-69134 Ecully cedex, France.*

Short headline : SUBHARMONIC RESPONSE OF A HERTZIAN CONTACT

34 pages in all,

including 15 figures.

* Corresponding author.    Tel.: +33-4-72-18-62-95; fax: +33-4-78-43-33-83.

E-mail address: joel.perret-liaudet@ec-lyon.fr

**Abstract**

In this paper, the response of a normally excited preloaded Hertzian contact is investigated in order to analyze the subharmonic resonance of order 2. The nonlinearity associated with contact losses is included. The method of multiple scales is used to obtain the non-trivial steady state solutions, their stability, and the frequency-response curves. To this end, a third order Taylor series of the elastic Hertzian contact force is introduced over the displacement interval where the system remains in contact. A classical time integration method is also used in conjunction with a shooting method to take into account losses of contact. The theoretical results show that the subharmonic resonance constitutes a precursor of dynamic responses characterised by loss of contact, and consequently, the resonance establishes over a wide frequency range. Finally, experimental validations are also presented in this paper. To this end, a specific test rig is used. It corresponds to a double sphere-plane contact preloaded by the weight of a moving mass. Experimental results show good agreements with theoretical ones.

**Nomenclature**

Latin characters

| | |
|---|---|
| *A* | amplitude of the subharmonic component of the response |
| *B* | intermediate variable |
| *c* | damping coefficient |
| $D_p$ | differentiation with respect to independent time variables |
| $D_{pq}$ | differentiation of order two with respect to independent time variables |
| *E* | Young's modulus |
| $f_0$ | linearized natural frequency |
| $F_{exp}$ | dimensionless experimental contact force |
| *F(q)* | dimensionless restoring contact force |
| *k* | Hertzian constant |
| *m* | rigid moving mass |
| *N* | static load |
| *q(τ)* | dimensionless displacement response |
| *R* | ball radius |
| *t* | time |
| $T_n$ | order n independent time variables |
| *z(t)* | displacement response |
| $z_s$ | static contact compression |

Greek characters

| | |
|---|---|
| *β* | intermediate variable |
| *γ* | intermediate variable |
| *ε* | small parameter |
| *ζ* | damping ratio |
| $\lambda$ | detuning parameter |
| $\nu$ | Poisson ratio |
| $\sigma$ | dimensionless level of the excitation force |
| $\sigma^*$ | first threshold of the dimensionless excitation force level |
| $\sigma^{**}$ | second threshold of the dimensionless excitation force level |
| $\tau$ | dimensionless time |
| $\phi$ | phase of the subharmonic component of the response |
| $\omega$ | circular frequency of the excitation force |
| $\bar{\omega}$ | dimensionless excitation circular frequency |
| $\Omega$ | linearized natural circular frequency |

Accents

| | |
|---|---|
| $\hat{\phantom{x}}$ | indicates O(1) variable |
| $\cdot$ | differentiation with respect to time t or dimensionless time $\tau$ |

# 1. Introduction

Hertzian contacts exist in many mechanical systems such as mechanisms and machines (gears, cam systems, rolling element bearings, to name a few). Under operating conditions, these contacts are often excited by dynamic normal forces superimposed on a mean static load. Under excessive excitation, contacts can exhibit undesirable vibroimpact responses, as a result of clearances introduced through manufacturing tolerances. The resulting dynamic behaviour is characterised by loss of contact and impacts, leading to excessive wear, surface damage and noise.

In a previous paper [1], the dynamic behaviour of an idealized preloaded and non-sliding dry Hertzian contact was studied under primary resonance conditions. To this end, an experimental test rig was built in order to investigate the resonance in detail, including vibroimpact responses. Theoretical results were also presented showing good agreement with the main characteristics of the primary resonance. In a companion paper [2], analysis was extended to the case of vibroimpact responses under Gaussian white random normal excitation.

The present work is concerned with the subharmonic response of the identical fundamental Hertzian contact under harmonic excitation. Several earlier papers discuss the dynamic response of Hertzian contacts [3–11], but to our knowledge, theoretical and experimental analyses of the 2-subharmonic resonance are rarely presented [12]. In this paper, the dynamic model studied is described in section 2. Theoretical results are presented in section 3, and finally experimental investigations and results are presented in section 4.

# 2. The dynamic model

## 2.1 Equation of motion

The system under study corresponds to the single-degree-of-freedom impact oscillator shown in Figure 1. A moving rigid mass *m* is kept in contact with a flat surface and loaded by a static normal force N. Assuming a Hertzian contact law, the non-linear restoring contact force is derived from material properties and contact geometry [13]. When the system is excited by a purely harmonic normal force superimposed on the static load, the equation of motion may be written as follows:

$$\begin{aligned} m\ddot{z} + c\dot{z} + k z^{3/2} &= N(1+\sigma \cos \omega t) \quad \text{for } z \geq 0 \\ m\ddot{z} + c\dot{z} + &= N(1+\sigma \cos \omega t) \quad \text{for } z < 0 \end{aligned} \quad (1)$$

where $z$ is the normal displacement of the rigid mass $m$ measured such that $z<0$ corresponds to loss of contact, $c$ is a damping coefficient, and $k$ is a constant given by the Hertzian theory. Furthermore, $\sigma$ controls the level of the excitation and $\omega$ denotes the excitation angular frequency.

When the excitation force is zero, the static contact compression $z_S$ is given by

$$z_S = \left(\frac{N}{k}\right)^{2/3} \quad (2)$$

we introduce the linearized contact natural frequency $\Omega$ and the damping ratio $\zeta$ given by

$$\Omega^2 = \left(\frac{3k}{2m}\right) z_S^{1/2} \quad (3)$$

$$\zeta = \frac{c}{2m\Omega} \quad (4)$$

and rescale Eq. (1) by defining

$$q = \frac{3(z-z_S)}{2z_S} \quad (5)$$

$$\tau = \Omega t. \quad (6)$$

The dimensionless equation of motion is obtained as follows [1]

$$\begin{aligned} \ddot{q} + 2\zeta \dot{q} + \left[1+\frac{2}{3}q\right]^{3/2} &= 1+\sigma \cos \overline{\omega} \tau \quad \text{for} \quad q \geq -\frac{3}{2} \\ \ddot{q} + 2\zeta \dot{q} &= 1+\sigma \cos \overline{\omega} \tau \quad \text{for} \quad q < -\frac{3}{2} \end{aligned} \quad (7)$$

In this equation, overdot indicates differentiation with respect to the dimensionless time $\tau$, and $\bar{\omega}$ is the dimensionless excitation circular frequency defined as follows

$$\bar{\omega} = \frac{\omega}{\Omega} \quad (8)$$

It should be noted that loss of contact now corresponds to the inequality

$$q \leq -\frac{3}{2} \quad (9)$$

*2.2 Approximation of the elastic contact force*

In order to use the analytical method of multiple scales, the restoring elastic contact force is approximated by expanding the non-linearity in a third-order Taylor series around the static load. In this way, both quadratic and cubic non-linearities appear naturally, and the approximate elastic restoring force is given as follows

$$\left[1 + \frac{2}{3}q\right]^{3/2} \approx 1 + q + \frac{1}{6}q^2 - \frac{1}{54}q^3 \quad (10)$$

Then, the final dimensionless equation appropriate for the use of the method of multiple scales is:

$$\ddot{q} + 2\zeta\dot{q} + q + \frac{1}{6}q^2 - \frac{1}{54}q^3 = \sigma \cos\bar{\omega}\tau \quad (11)$$

This equation remains valid if the following inequality is satisfied

$$q \geq -1.348 \quad (12)$$

This inequality guarantees positive values for the approximate restoring force. In other words, this inequality corresponds to the loss of contact condition for the approximate system. Furthermore, the absolute difference between the actual and the approximate elastic restoring force is less than 0.01 in the range $-1 < q < 1$. Comparison is shown in Figure 2.

**3. The theoretical response to the order-2 subharmonic excitation**

*3.1. Multiple scales method*

Initially, the method of multiple scales is used [14-15]. To this end, the ordering

$$q = \varepsilon \hat{q}, \zeta = \varepsilon \hat{\zeta}, \sigma = \varepsilon \hat{\sigma} \qquad (13)$$

is assumed in Eq. (11) where circumflexes indicate O(1) variables. As a result, the excitation appears at the same order as the free response (i.e. at the order $\varepsilon^0$), and the damping appears at the same time as the quadratic non-linearity (i.e. at the order $\varepsilon^1$). Then Eq. (11) becomes, omitting circumflexes in Eq. (13),

$$\ddot{q} + q + \varepsilon(2\zeta\dot{q} + \frac{1}{6}q^2) - \frac{\varepsilon^2}{54}q^3 = \sigma\cos\bar{\omega}\tau \qquad (14)$$

By introducing new independent time variables $T_n = \varepsilon^n t$ (with n = 0, 1, 2), expanding $q$ in power series of $\varepsilon$,

$$q = \sum_{n=0}^{n=2} \varepsilon^n q_n(T_0, T_1, T_2) \qquad (15)$$

and equating coefficients of like powers of $\varepsilon$, one obtains the following system of perturbation equations at the first three $\varepsilon$ orders [15]:

$$D_{00}q_0 + q_0 = \sigma\cos\bar{\omega}T_0$$

$$D_{00}q_1 + q_1 = -(2D_{01} + 2\zeta D_0)q_0 - \frac{1}{6}q_0^2$$

$$D_{00}q_2 + q_2 = -(2D_{01} + 2\zeta D_0)q_1 - (2D_{02} + D_{11} + 2\zeta D_1)q_0 - \frac{1}{3}q_0 q_1 + \frac{1}{54}q_0^3 \quad (16\text{ a-c})$$

where

$$D_p = \frac{\partial}{\partial T_p}, \quad D_{pq} = \frac{\partial^2}{\partial T_p \partial T_q}. \qquad (17)$$

To analyse the subharmonic resonance of order 2, one expresses the nearness of the external excitation frequency to twice the linearized natural frequency by introducing the detuning parameter $\lambda$ defined according to

$$\bar{\omega} = 2 + \varepsilon\lambda \qquad (18)$$

Solving system (16), eliminating secular terms and retaining steady state solutions leads to the non-trivial steady state response at the $\varepsilon$ order as follows

$$q(\tau) = -\frac{1}{3}(B^2 + \frac{A^2}{4}) + A\cos(\frac{\overline{\omega}\tau - \phi}{2})$$
$$+ 2B\cos(\overline{\omega}\tau) + \frac{A^2}{36}\cos(\overline{\omega}\tau - \phi) + \frac{4\zeta\overline{\omega}B}{1-\overline{\omega}^2}\sin(\overline{\omega}\tau) \quad (19)$$
$$+ \frac{AB}{3\overline{\omega}(\overline{\omega}+2)}\cos(\frac{3\overline{\omega}\tau - \phi}{2}) + \frac{B^2}{3(4\overline{\omega}^2 - 1)}\cos(2\overline{\omega}\tau)$$

where

$$B = \frac{\sigma}{2(1-\overline{\omega}^2)} \quad (20)$$

In Eq. (19) the phase $\phi$ is given by

$$\tan\phi = \frac{(\overline{\omega}^2 - 1)(\frac{\lambda}{2} - 1) - \overline{\omega}(\lambda - \gamma + \frac{A^2}{27})}{[(1-\overline{\omega}^2)(\frac{\lambda}{2} - 1)(\lambda - \gamma + \frac{A^2}{27})/2\zeta] - \zeta\overline{\omega}} \quad (21)$$

where

$$\gamma = [-\frac{7}{36} + \frac{1}{9}(\overline{\omega}^2 + 2\overline{\omega})]B^2 - \zeta^2 \quad (22)$$

Also, the amplitude $A$ of the component at one half the excitation frequency is determined by the following frequency response equation

$$\frac{A^2}{27} = (\gamma - \lambda) \pm (\beta^2\sigma^2 - 4\zeta^2)^{1/2} \quad (23)$$

where

$$\beta = \frac{1}{\overline{\omega}^2 - 1}[\frac{1}{36}(1 - \frac{\lambda}{2})^2 + (\frac{\zeta\overline{\omega}}{3\overline{\omega}^2 - 3})^2]^{1/2} \quad (24)$$

Next, it is easy to precise the regions of existence of parameter space where subharmonic responses exist, by determining the real roots of Eq. (23). The stability of the solutions can also be easily obtained [15]. Introducing the two critical values defined by

$$\sigma_1 = \frac{2\zeta}{|\beta|}$$

$$\sigma_2 = \frac{[(\lambda-\gamma)^2 + 4\zeta^2]^{1/2}}{|\beta|} \qquad (25\ a\text{-}b)$$

one can distinguish the three following regions, illustrated in Figure 3 for the case $\zeta = 0.005$. In region I defined by $\sigma > \sigma_2$, the trivial response (or 1T-periodic response) is unstable and subharmonic responses are always excited. In region II defined by $\sigma_1 < \sigma < \sigma_2$ and $\bar{\omega} < 2$, the trivial response is stable and it coexists with two subharmonic responses, one of which stable, the other unstable. For this case, the stable subharmonic response can be excited depending on the initial conditions. In region III, when $\sigma < \sigma_1$ and $\bar{\omega} < 2$, or when $\sigma < \sigma_2$ and $\bar{\omega} > 2$, the trivial 1T-periodic response is always stable and subharmonic response can never be excited. In other words, the curve $\sigma = \sigma_2$ corresponds to the subharmonic bifurcation which is subcritical when $\bar{\omega} < 2$ and critical when $\bar{\omega} > 2$. Following these results, one obtains a critical excitation level $\sigma^*$ which can be viewed as a threshold beyond which the subharmonic resonance is always excited. This value is given according to

$$\sigma^* = \frac{36\zeta}{1 + 16\zeta^2/9} \qquad (26)$$

and if $\zeta \ll 1$:

$$\sigma^* \approx 36\zeta \qquad (27)$$

Beyond this threshold, the subharmonic resonance exhibits a softening behaviour shown in Figure 4. This figure displays the frequency response curve obtained for $\sigma = 0.5$ and $\zeta = 0.01$. This result is in agreement with the behaviour of the system under primary resonance conditions in the sense that the frequency response curve is bent to low frequency (softening behaviour) [1].

Figure 5 shows the minimum value of the displacement response when subharmonic resonance occurs. By considering the approximate system, Eq. (11), one can see that the 2-subharmonic response leads to contact loss. In fact, we have always observed this behaviour which constitutes a general trend of the system. The question can now be stated as follows: does the 2-superharmonic response always lead to loss of contact for the original dynamic model defined by Eq. (7)? This question requires a more detailed analysis of superharmonic resonance condition. We therefore supplement this theoretical study by a numerical investigation described as below.

*3.2. Continuation method*

In order to achieve dynamic responses including loss of contact, a classical numerical time integration explicit scheme is used, i.e. the central difference method. The interest of using an explicit scheme is the estimation of the non linear restoring force which is not required at each time step by a non-linear solver, like the Newton-Raphson method. The counterpart is to use a sufficiently small time step examining both the period of the response and the linearized natural frequency of the system. In our simulations, we have imposed a time step 10000 times less than the period of the response. Further, a specific computing method devoted to nonlinear problems is used, namely the shooting method combined with a continuation technique. For details of these known methods, see for example references [16, 17]. In our case, the shooting method is similar to tracking fixed points on a Poincaré map. To this end, we have chosen a Poincaré section equivalent to a stroboscopic section at a period 1 or 2 times the period of the external excitation. Finally, the required Jacobian matrices attached to the fixed points are evaluated via a numerical rule. Actually, these can not be analytically determinate because of the non regular characteristic of the non-linear restoring force. One can also notice that the stability of the response and the kind of bifurcations can be deduced from the eigenvalues of the Jacobian matrices.

Figure 6 shows a typical frequency-response curve which is obtained for the same values as used in Figures 4 and 5 ($\sigma = 0.5$ and $\zeta = 1$ %). As we can see in the upper part (Figure 6(a) ), the subharmonic resonance is very strong as it is established over a wide frequency range. Actually, the downward jump frequency appears to be very low ($\varpi \approx 0.475$). As one observes in the lower part (Figure 6(b) ) which is the frequency-response curve detailed around $\varpi \approx 2$, we confirm the multiple scales method result, i.e. the fact that the subharmonic resonance initiates vibroimpact responses. This is of great importance in a practical point of view when vibroimpacts lead to excessive wear, surface damage and excessive noise.

In order to describe the bifurcation set obtained by the used numerical shooting method, we have tracked the 1T-periodic response for different values of $\sigma$ and identified frequencies for which subharmonic bifurcations occur. This can be done by computing the multipliers or eigenvalues of the Monodromy matrix related to the associated fixed point. Remind that subharmonic bifurcation occurs when one of the multipliers leaves the unit circle in the complex plane by the $-1$ value. Comparison of the results with those obtained by the multiple scales method is given in Figure 7. As we can see, very good agreement between the two approaches occurs.

In our simulations, we have found that the loss of contact nonlinearity is also at the origin of subharmonic responses which then exhibite impacts. To illustrate this result, Figure 8 displays an example of frequency response curve obtained with an excitation amplitude $\sigma$ quite lower than the threshold value $\sigma^*$ defined by Eq. (26). As we can see, these subharmonic responses take place on an isola, i.e. a loop in the bifurcation set, delimited by a couple of saddle-node bifurcations. This isola establishes itself in the interval $1 < \overline{\omega} < 2$, which is coherent with the softening character of the loss of contact nonlinearity. Figure 9 displays the saddle-node bifurcation curve in the $\overline{\omega} - \sigma$ plane which circumscribes the region where isola takes place. In this Figure, the excitation amplitude threshold quoted $\sigma^{**}$ ($\sigma^{**} \approx 0.165$) corresponds to the isola formation at a frequency close to 1.5. It is found quite lower than the preceding one $\sigma^*$ defined by Eq. (26). Up to this threshold, we observe that isola rapidly grows when the excitation amplitude increases until it meets the preceding flip bifurcation close to $\bar{\omega} = 2$ and $\sigma = \sigma^*$. For these conditions, one can assume the existence of an unstable transcritical bifurcation. Actually, this last takes place very close to the conditions $\sigma = 0.336$ and $\zeta = 1\ \%$, as illustrated in Figure 10.

The result is of great importance as it proves that the subharmonic resonance initiated by the Hertzian nonlinearity almost always induces vibroimpact responses in a wide frequency range.

## 4. Experimental validation

### 4.1. Test rig

The main goals of our experimental study are to confirm the excitation level threshold value defined by Eq. (26) and to confirm the occurrence of vibroimpact responses initiated by subharmonic resonance conditions. In order to perform these experimental validations, we have used a test rig similar to that described in references [1-2]. The used test rig is depicted in Figure 11. It consists on a 25.4 mm diameter SAE 52100 steel ball preloaded between two horizontal SAE 52100 steel flat surfaces. The first one is fixed to a heavy rigid frame and the second one is rigidly fixed to a vertically moving cylinder. Compliance of a rough and weakly loaded contact obtained experimentally can be different from the theoretical compliance supplied by the Hertz equation. So, to take into account this problem, planes were ground to obtain roughness Ra < 0,4 µm. Ball roughness is also weak (Ra < 0.03 µm). Then, as we will see, asperities will be quite smaller than contact deflection and contact area. The double sphere-plane dry contact is preloaded by a static normal load $N = 69.7$ N, which corresponds to the weight of the moving cylinder ($m = 7.1$ kg). By assuming identical mechanical properties for the ball and the discs, the constant $k$ of the restoring elastic force expression is deduced from the double sphere-plane Hertzian contact as follows [13]:

$$k = \frac{E\sqrt{R}}{3\sqrt{2}(1-\nu^2)} \tag{28}$$

where $E$ is the Young's modulus (210 GPa), $\nu$ is the Poisson ratio (0.29) and $R$ is the ball radius (12.7 mm). Then, the main theoretical characteristics of the experimental system are:

$$k = 6.1 \; 10^9 \; \text{Nm}^{-3/2}$$

$$z_S = 5.1 \, \mu\text{m}$$

$$f_0 = \frac{\Omega}{2\pi} = 271 \, \text{Hz} \tag{29 a-c}$$

Contact is normally excited by a suspended vibration shaker. Harmonic normal force is applied to the moving cylinder and superimposed on the static load. Excitation force and normal force transmitted to the frame through the contact are measured by piezoelectric force transducers. Classical charge amplifiers are used for all responses.

*4.2. Measured natural frequency and damping ratio*

Linearized contact frequency ($f_0 = 270.6$ Hz) and equivalent viscous damping ratio (0.5 %) are measured from the almost linear contact dynamic behaviour under very low external input amplitude [1, 6]. The experimental natural frequency is close to the theoretical one since the relative error is less than 0.15 %, and the damping ratio value is coherent with preceding studies [1-2, 6].

*4.3. Experimental subharmonic resonance*

For this set of experimental results, it is important to say that good repeatability was always observed. Experimental subharmonic resonance is exhibited for external input amplitude $\sigma$ up to 20 %. This value can be considered as the critical excitation level $\sigma^*$ theoretically defined by Eq. (26) or Eq. (27). By considering the measured viscous damping ratio (0.5 %), the experimental ratio, that is $(\sigma^*/\zeta)_{exp} \approx 40$, appears to be in a good agreement with the theoretical one, that is $(\sigma^*/\zeta)_{th} = 36$. Figure 12 displays a typical result obtained for an external input amplitude $\sigma \approx 30\%$. It consists on the frequency response curves for the two first harmonics of the transmitted force, quoted $H_1$ and $H_2$ and respectively associated to the external input frequency $\overline{\omega}$ and to its first harmonic $2\overline{\omega}$. The jump discontinuities associated to the 2-subharmonic resonance are clearly observed. In particular, tracking the $H_2$ component allows to identify one of the two flip bifurcations. The subharmonic resonance leads to intermittent loss of contact. This source of nonlinearity noticeably dominates the dynamic behaviour of the system. Actually, Figure 13 shows that it strongly bends the frequency response curve to low frequency with a downward jump frequency less than the linearized contact natural frequency. So, for decreasing external input frequency, dynamic vibroimpact response is established over a wide frequency range, at least from $2\Omega$ to $\Omega$. In fact, this result is obtained for all the experimental conditions, even for an external input amplitude close to the amplitude threshold $\sigma^* \approx 20\%$.

Finally, time histories of the dimensionless normal force observed in subharmonic resonance conditions are displayed in Figures 14 and 15. Notice that the contact force $F_{exp}$ is dimensionalized after centered it regards to the static load as follows:

$$F_{exp} = \frac{F(t) - N}{N} \tag{30}$$

So, contact is lost when the dimensionless normal contact force is under the -1 value.

These are obtained in the same conditions that those imposed in Figure 12, i.e. with an external input amplitude $\sigma \approx 30\%$. The flip bifurcation is clearly identified when the period of the normal force is twice of its of the external excitation. Time histories show also the hardening behaviour of the dynamic system in compression and its softening behaviour in extension. Finally, plates shown in Figure 15 correspond to the flight duration when loss of contact occurs.

## 5. Conclusion

In this study, the dynamic response of an impacting Hertzian contact subjected to an order-2 subharmonic excitation is analyzed. The critical excitation level beyond which subharmonic resonance always occurs is theoretically predicted and experimentally confirmed. Beyond this excitation level, response is quickly characterized by vibroimpacts and the dynamic behaviour is mainly governed by loss of contact nonlinearity. Although contact stiffness is modelled by a linear law in most o classical impact models, we showed that the nonlinearity inherent in the Hertzian law or more general law, contact laws cannot be ignored if one wishes to predict vibroimpact response, because it is this nonlinearity which initiates the subharmonic resonance.

**Captions for Figures**

Figure 1. Dynamic model of the single-degree-of-freedom impact oscillator.

Figure 2. The dimensionless restoring contact force model (thick line) and its approximate form defined by Eq. (10) (thin line).

Figure 3. Bifurcation set in the $\bar{\omega} - \sigma$ plane for $\zeta = 0.5\%$.

Figure 4. Frequency response curve $A(\bar{\omega})$ exhibiting the subharmonic resonance ($\sigma = 0.5$, $\zeta = 1\%$).

Figure 5. Frequency response curve Min($q$) exhibiting loss of contact under the expected value $q = -1.348$, see Eq. (12), and for $\sigma = 0.5$ and $\zeta = 1\%$.

Figure 6. Frequency peak to peak response curve of $q$ obtained by the shooting method for $\sigma = 0.5$ and $\zeta = 1\%$ : (a) the complete response curve; (b) the detail of the response curve around $\bar{\omega} = 2$ (I: 1T-stable responses; II: 2T-stable responses without loss of contact; III: 2T-stable responses with loss of contact; IV: 2T-unstable responses without loss of contact; V: 2T-unstable responses with loss of contact).

Figure 7. Bifurcation set (Flip bifurcation) in the $\bar{\omega} - \sigma$ plane for $\zeta = 0.5\%$ obtained by the multiple scales method (─) and the continuation method (□).

Figure 8. Frequency peak to peak response curve of $q$ exhibiting the isola of the 2T-periodic response and obtained by the shooting method for $\sigma = 0.2$ and $\zeta = 1\%$.

Figure 9. Bifurcation set in the $\bar{\omega} - \sigma$ plane for $\zeta = 1\%$. Thick line represents the saddle node bifurcation for the 2T response associated with the isola. In this figure is also reported the bifurcation set associated with the order-2 subharmonic resonance induced by the Hertzian law (thin line).

Figure 10. Frequency peak to peak response curve of $q$ exhibiting the unstable transcritical bifurcation ($\sigma = 0.336$ and $\zeta = 1\%$). Thick line : stable response; thin line unstable response.

Figure 11. Photo of the used dynamic test rig.

Figure 12. Experimental H1 (a) and H2 (b) harmonic response curves versus the dimensionless excitation circular frequency $\bar{\omega}$ (with $\bar{\omega} > 1.9$) and obtained for a dimensionless excitation force $\sigma \approx 30\%$.

Figure 13. Experimental H1 (a) and H2 (b) harmonic response curves versus the dimensionless excitation circular frequency $\bar{\omega}$ (with $\bar{\omega} > 0.9$) and obtained for a dimensionless excitation force $\sigma \approx 30\%$.

Figure 14. Time histories of the dimensionless normal force $F_{exp}$ versus the dimensionless time ($\bar{\omega} \tau / 2 \pi$) for $\bar{\omega} = 2.023$ (a), 2.020 (b) and 2.014 (c) and obtained for a dimensionless excitation force $\sigma \approx 30\%$.

Figure 15. Time histories of the dimensionless normal force $F_{exp}$ versus the dimensionless time ($\bar{\omega}\,\tau/2\pi$) for $\bar{\omega}$ = 1.996 (a), 1.914 (b) and 1.848 (c) obtained for a dimensionless excitation force $\sigma \approx 30\,\%$.

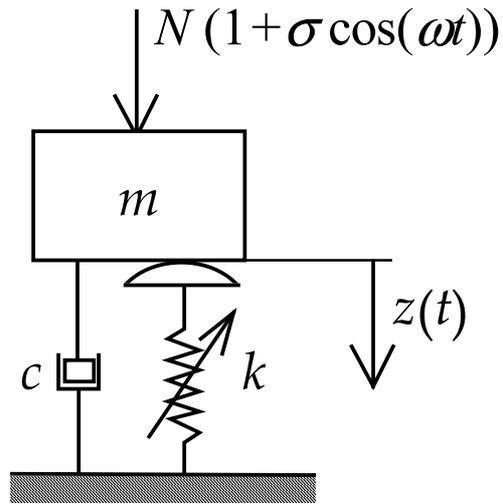

Figure 1. Dynamic model of the single-degree-of-freedom impact oscillator.

J. Perret-Liaudet and E. Rigaud.

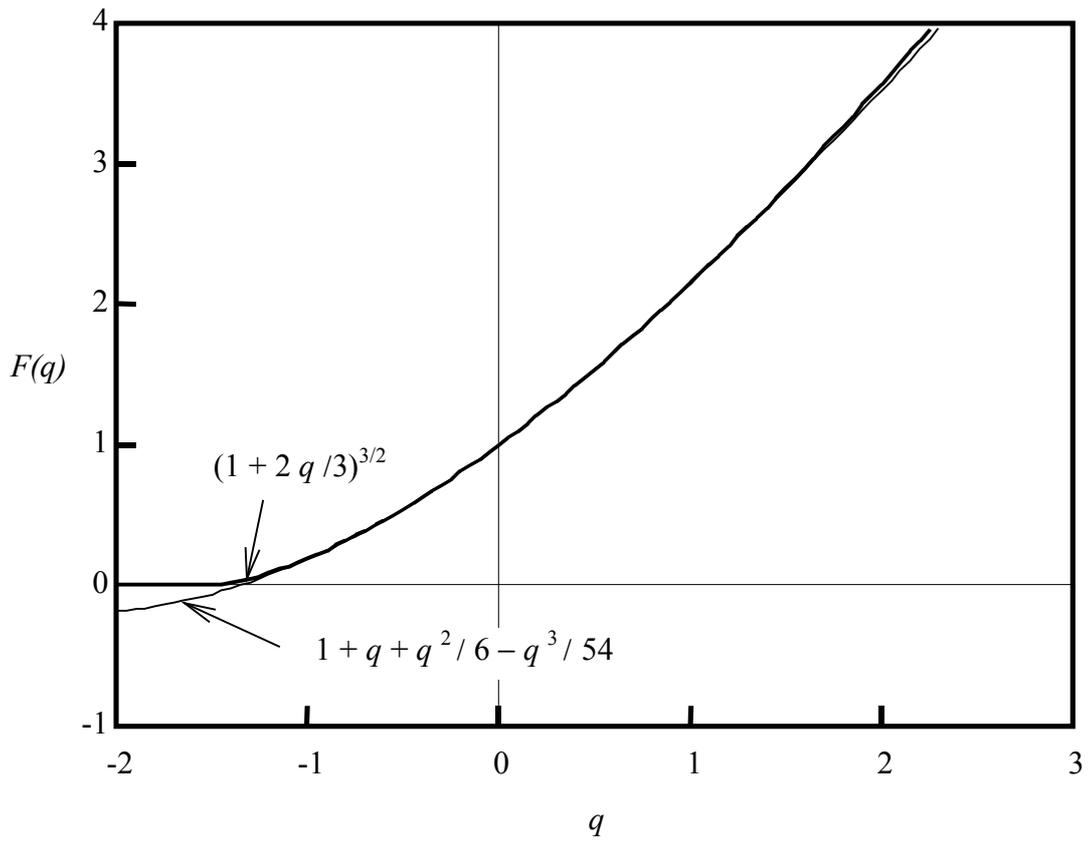

Figure 2. The dimensionless restoring contact force model (thick line) and its approximate form defined by Eq. (10) (thin line).

J. PERRET-LIAUDET and E. RIGAUD.

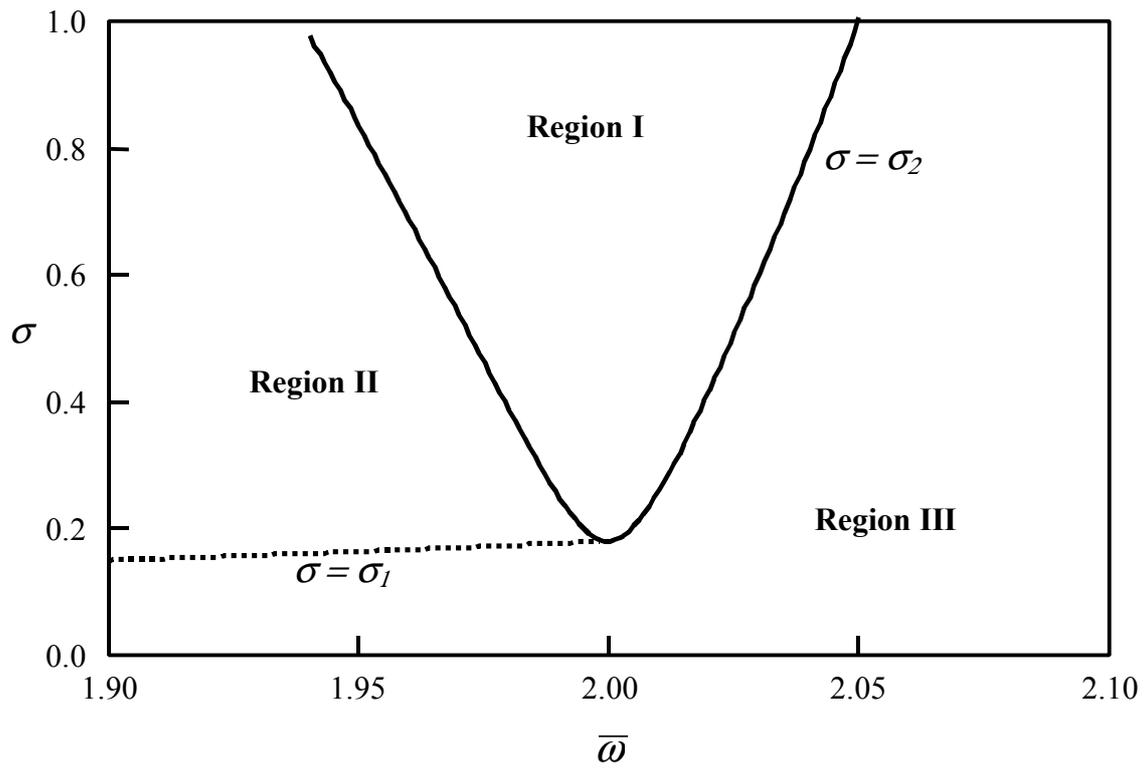

Figure 3. Bifurcation set in the $\bar{\omega} - \sigma$ plane for $\zeta = 0.5\%$.

J. PERRET-LIAUDET and E. RIGAUD.

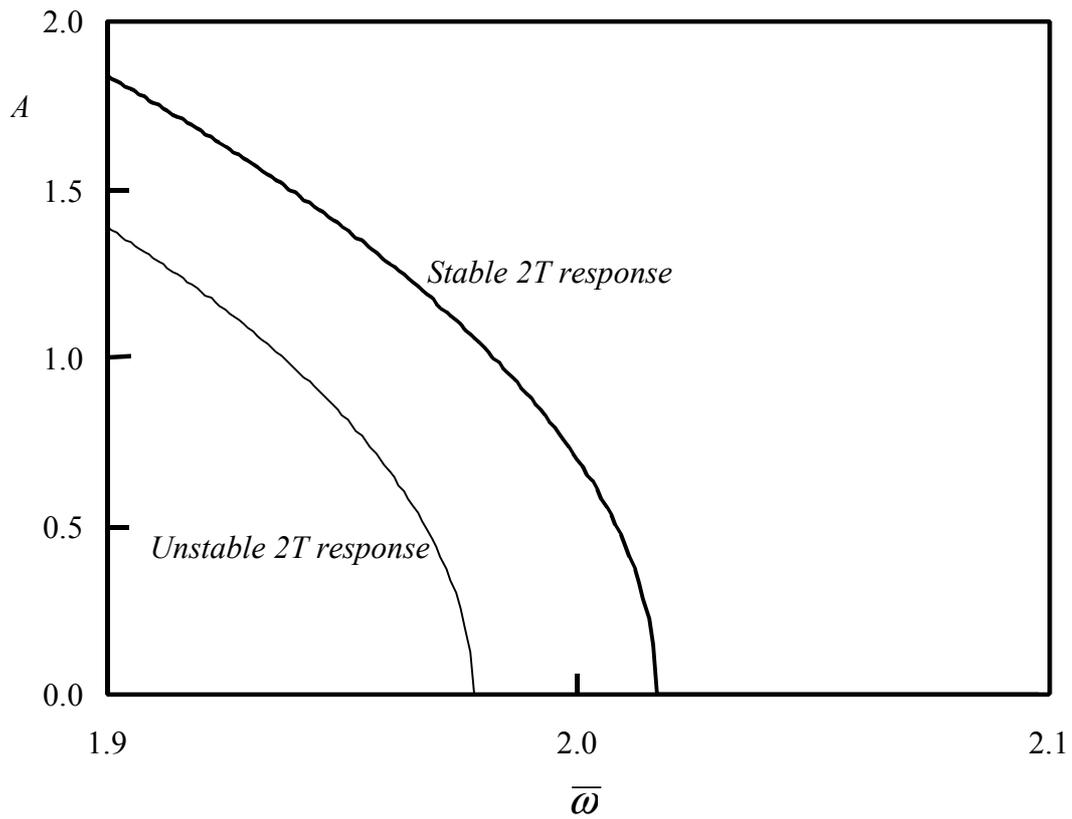

Figure 4. Frequency response curve $A(\bar{\omega})$ exhibiting the subharmonic resonance ($\sigma = 0.5$, $\zeta = 1\ \%$).

J. PERRET-LIAUDET and E. RIGAUD.

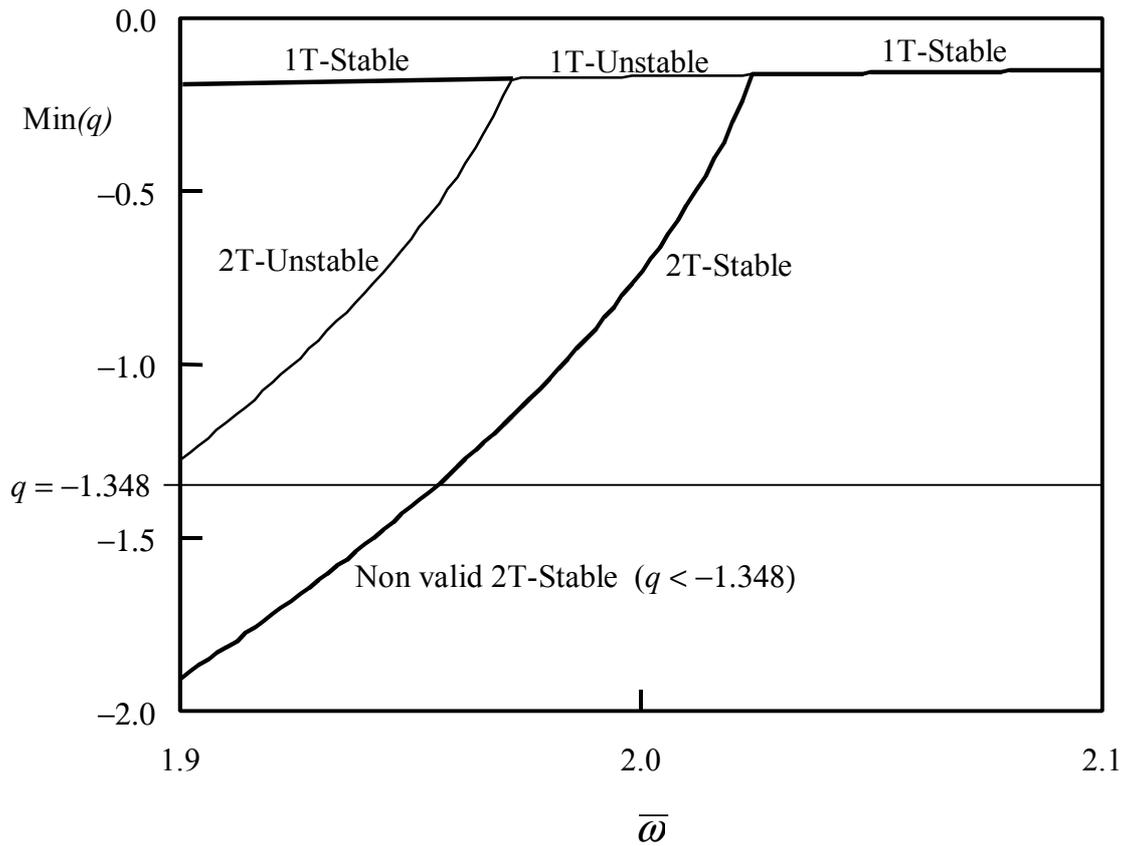

Figure 5. Frequency response curve Min($q$) exhibiting loss of contact under the expected value $q = -1.348$, see Eq. (12), and for $\sigma = 0.5$ and $\zeta = 1\%$.

J. PERRET-LIAUDET and E. RIGAUD.

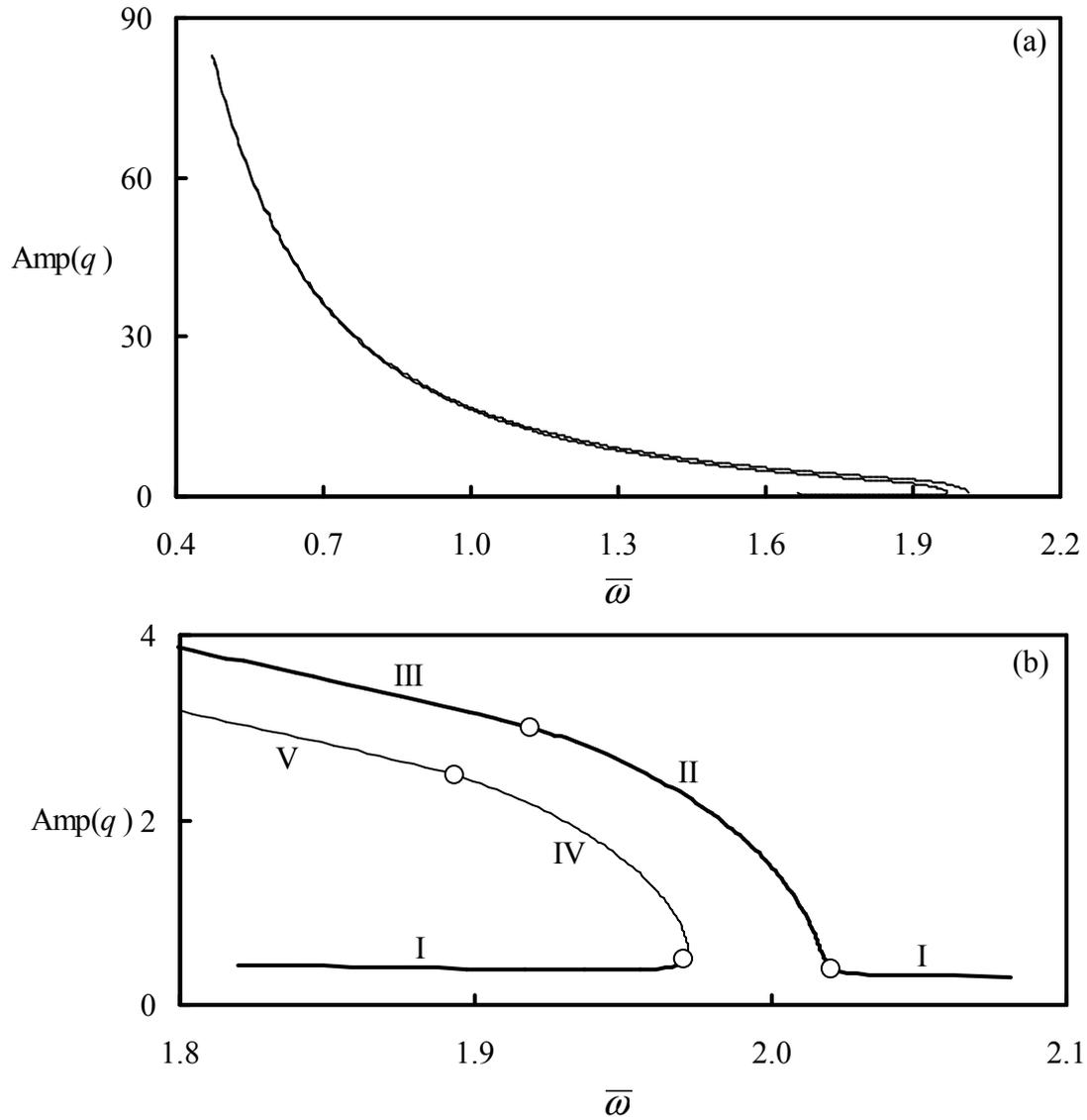

Figure 6. Frequency peak to peak response curve of $q$ obtained by the shooting method for $\sigma = 0.5$ and $\zeta = 1\%$ : (a) the complete response curve; (b) the detail of the response curve around $\bar{\omega} = 2$ (I: 1T-stable responses; II: 2T-stable responses without loss of contact; III: 2T-stable responses with loss of contact; IV: 2T-unstable responses without loss of contact; V: 2T-unstable responses with loss of contact).

J. PERRET-LIAUDET and E. RIGAUD.

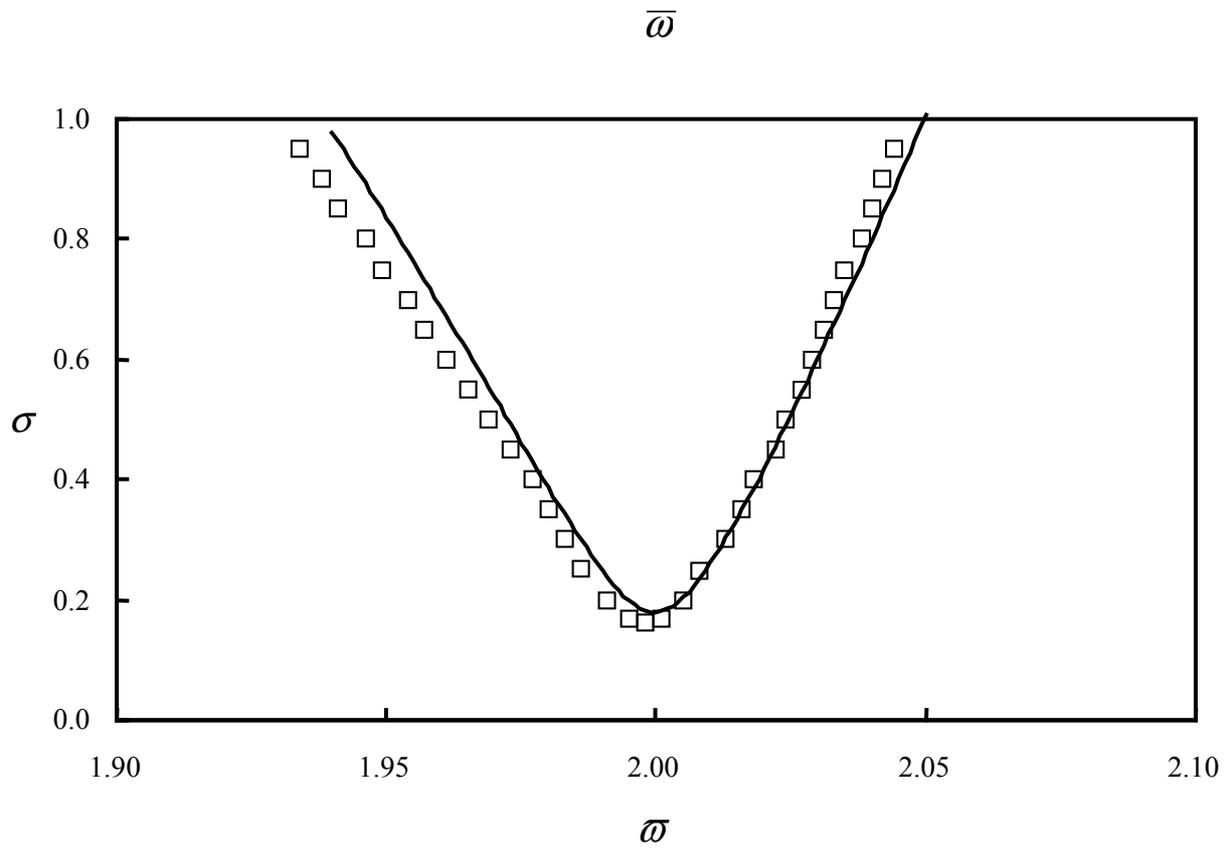

Figure 7. Bifurcation set (Flip bifurcation) in the $\bar{\omega} - \sigma$ plane for $\zeta = 0.5\%$ obtained by the multiple scales method (—) and the continuation method (□).

J. Perret-Liaudet and E. Rigaud.

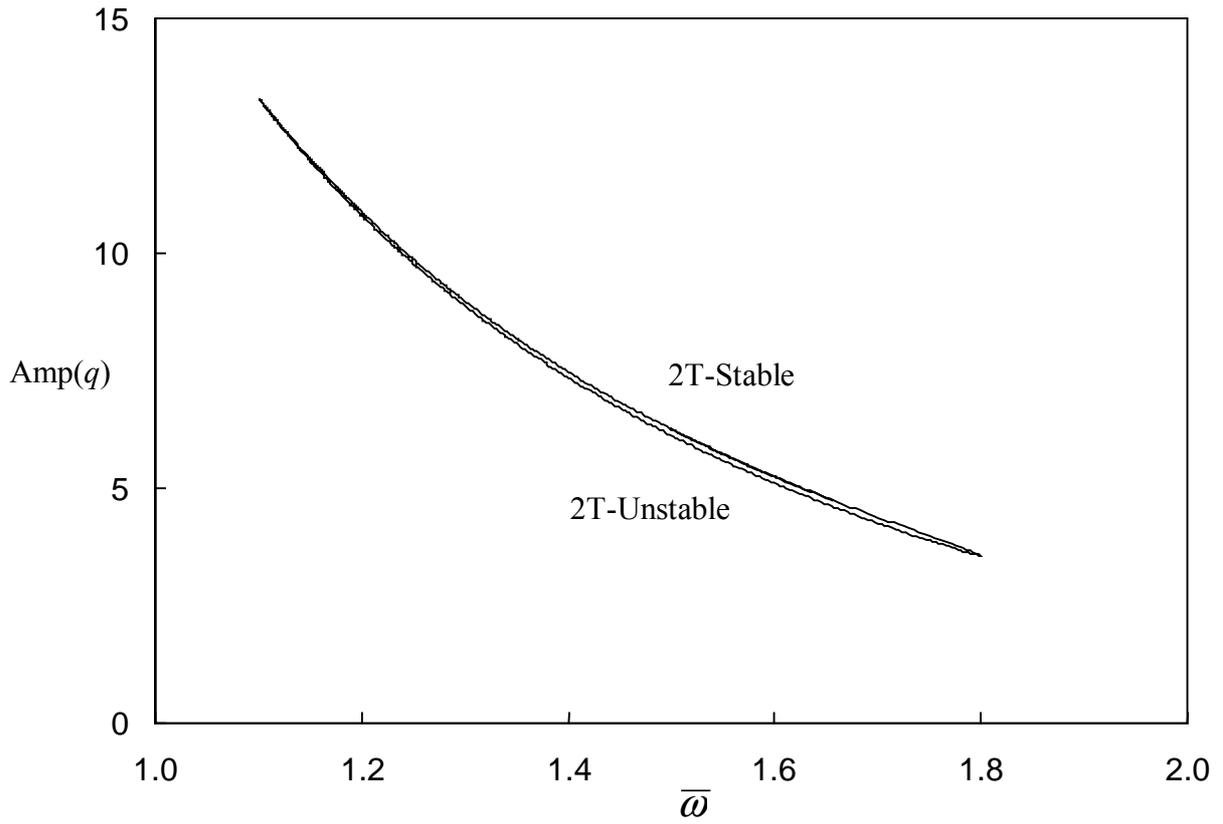

Figure 8. Frequency peak to peak response curve of $q$ exhibiting the isola of the 2T-periodic response and obtained by the shooting method for $\sigma = 0.2$ and $\zeta = 1\ \%$.

J. PERRET-LIAUDET and E. RIGAUD.

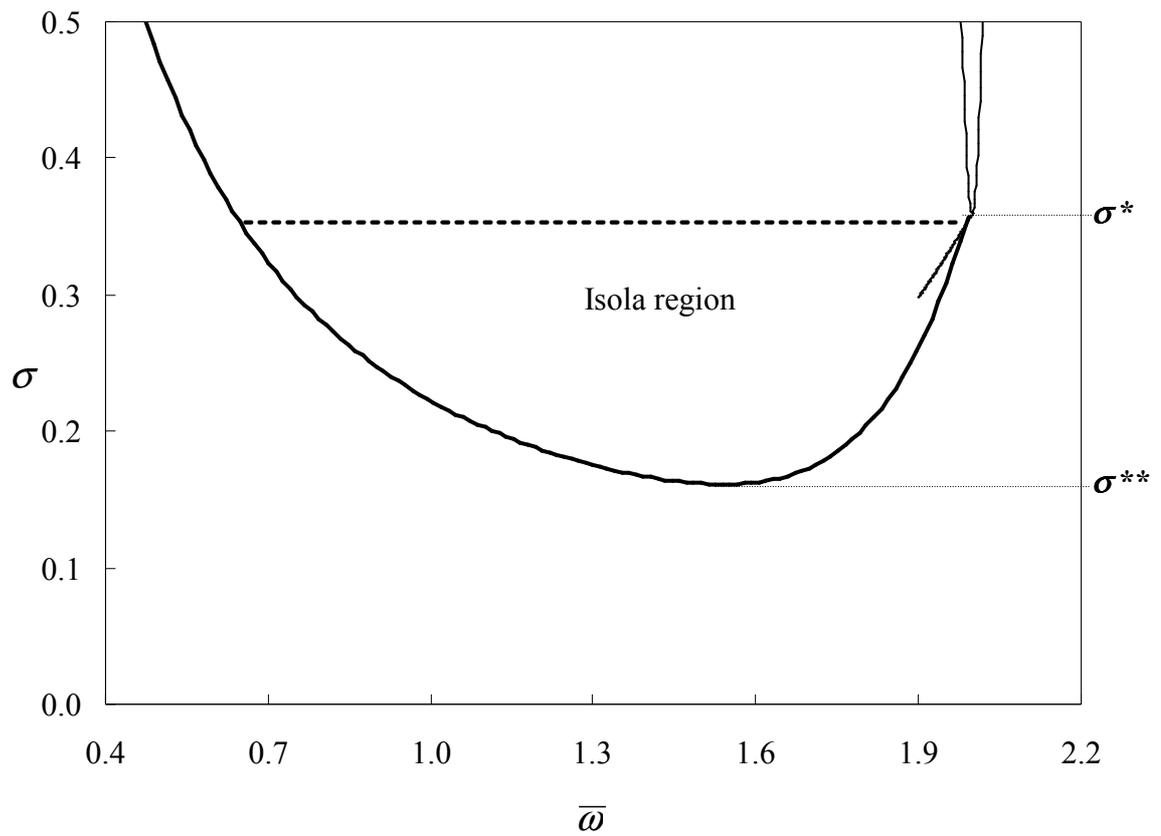

Figure 9. Bifurcation set in the $\bar{\omega} - \sigma$ plane for $\zeta = 1\ \%$. Thick line represents the saddle node bifurcation for the 2T response associated with the isola. In this figure is also reported the bifurcation set associated with the order-2 subharmonic resonance induced by the Hertzian law (thin line).

J. PERRET-LIAUDET and E. RIGAUD.

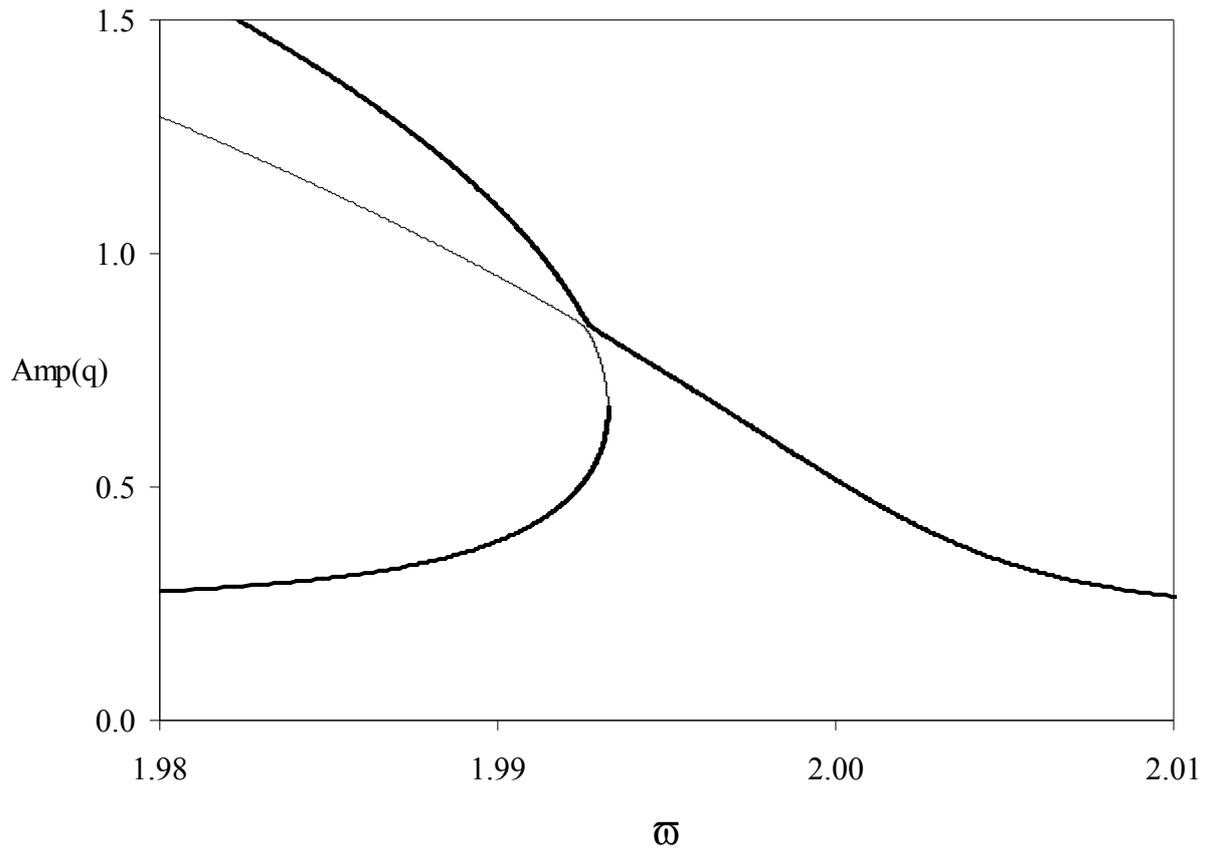

Figure 10. Frequency peak to peak response curve of $q$ exhibiting the unstable transcritical bifurcation ($\sigma = 0.336$ and $\zeta = 1\ \%$). Thick line : stable response; thin line unstable response.

J. PERRET-LIAUDET and E. RIGAUD.

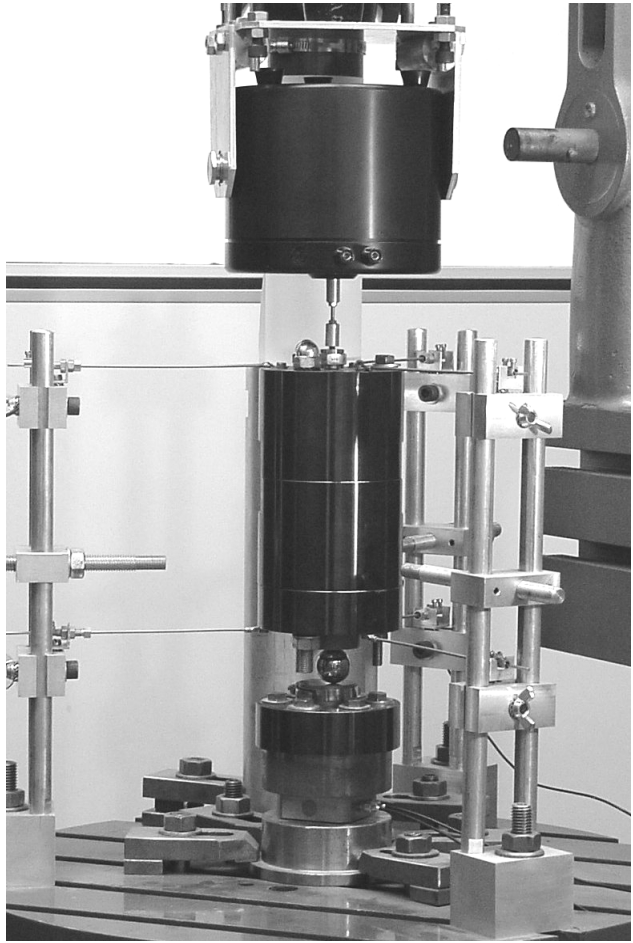

Figure 11. Photo of the dynamic test rig.

J. Perret-Liaudet and E. Rigaud.

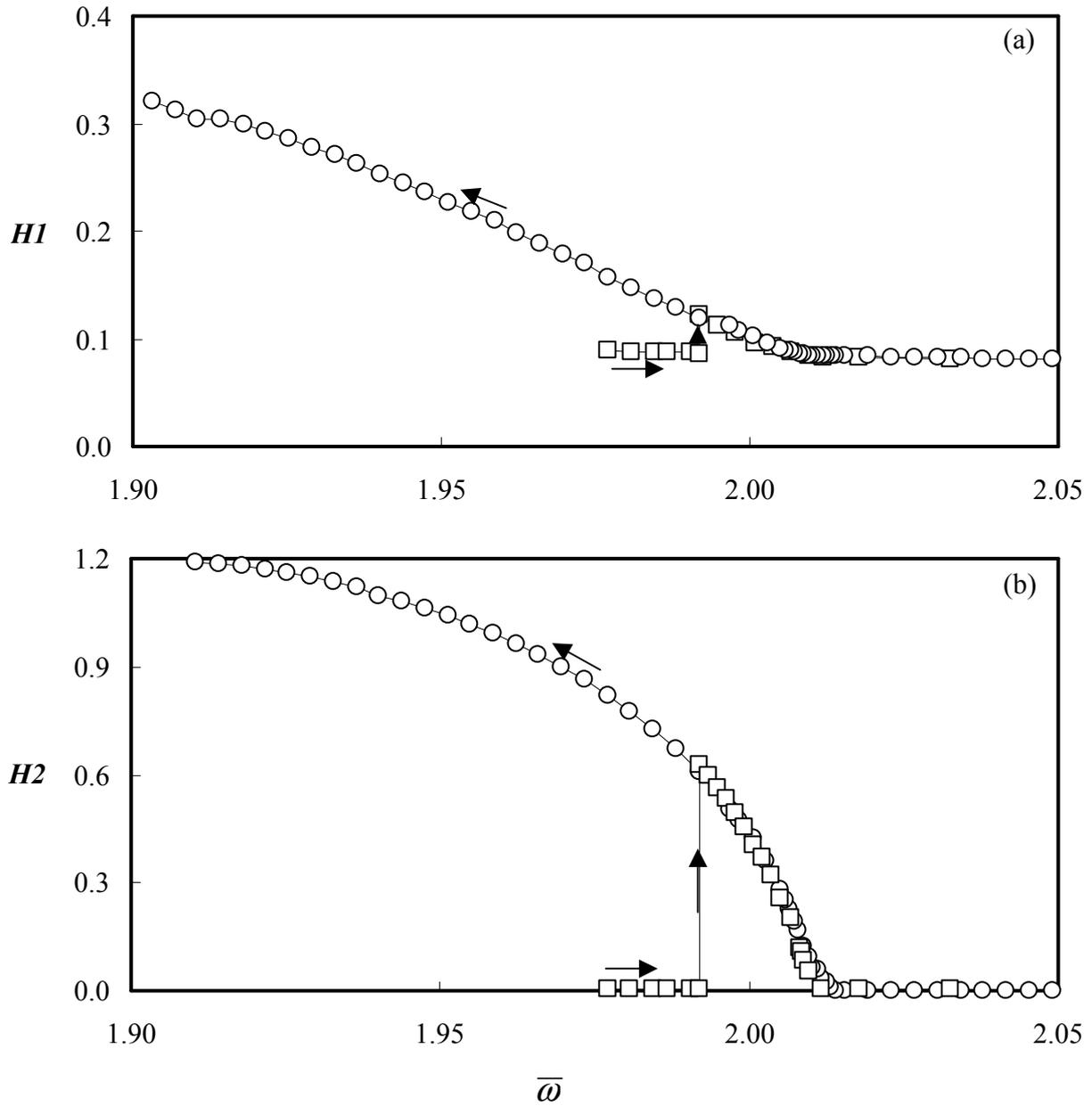

Figure 12. Experimental H1 (a) and H2 (b) harmonic response curves versus the dimensionless excitation circular frequency $\bar{\omega}$ (with $\bar{\omega} > 1.9$) and obtained for a dimensionless excitation force $\sigma \approx 30\ \%$.

J. PERRET-LIAUDET and E. RIGAUD.

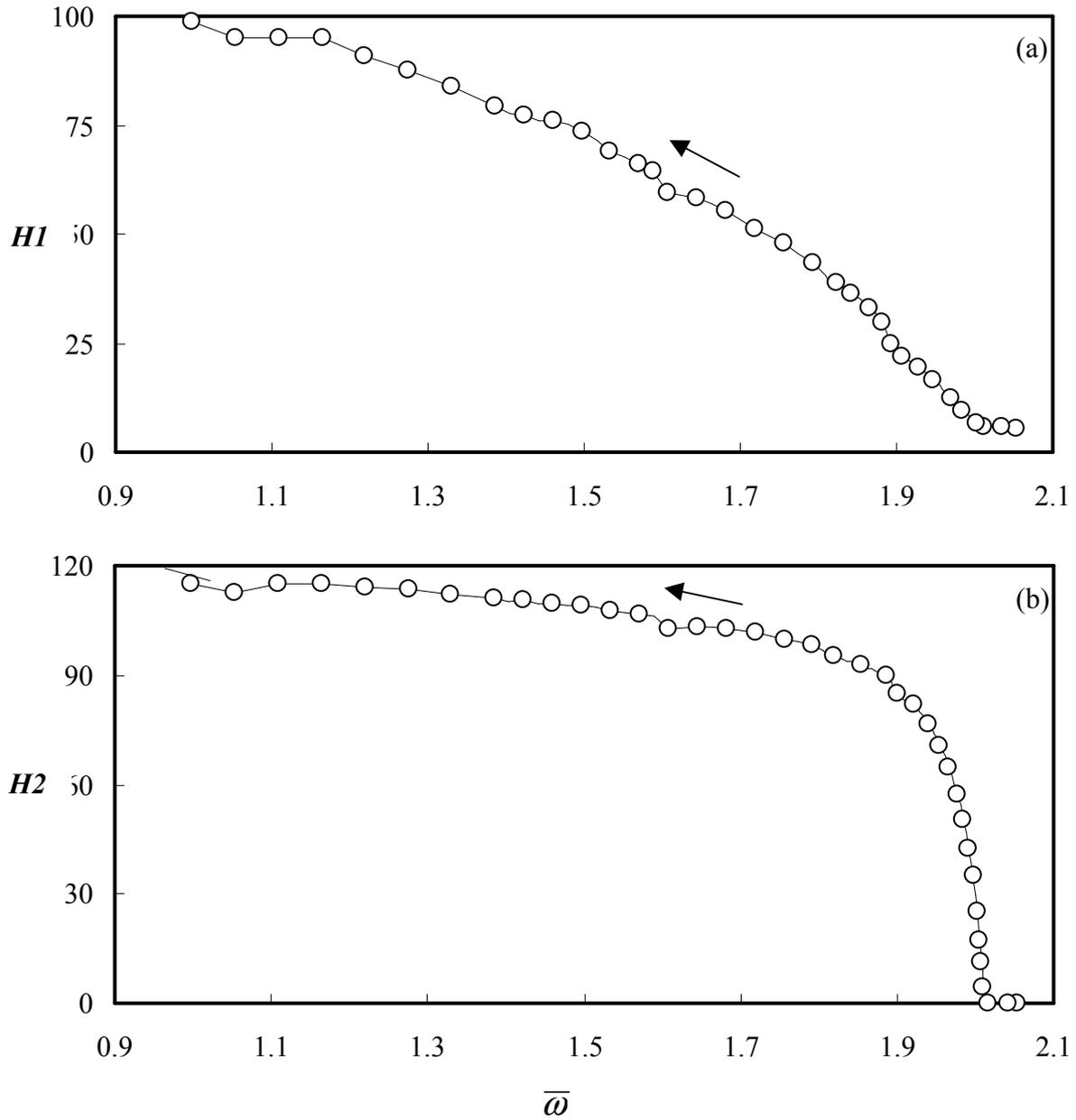

Figure 13. Experimental H1 (a) and H2 (b) harmonic response curves versus the dimensionless excitation circular frequency $\overline{\omega}$ (with $\overline{\omega}$ >0.9) and obtained for a dimensionless excitation force $\sigma \approx 30$ %.

J. PERRET-LIAUDET and E. RIGAUD.

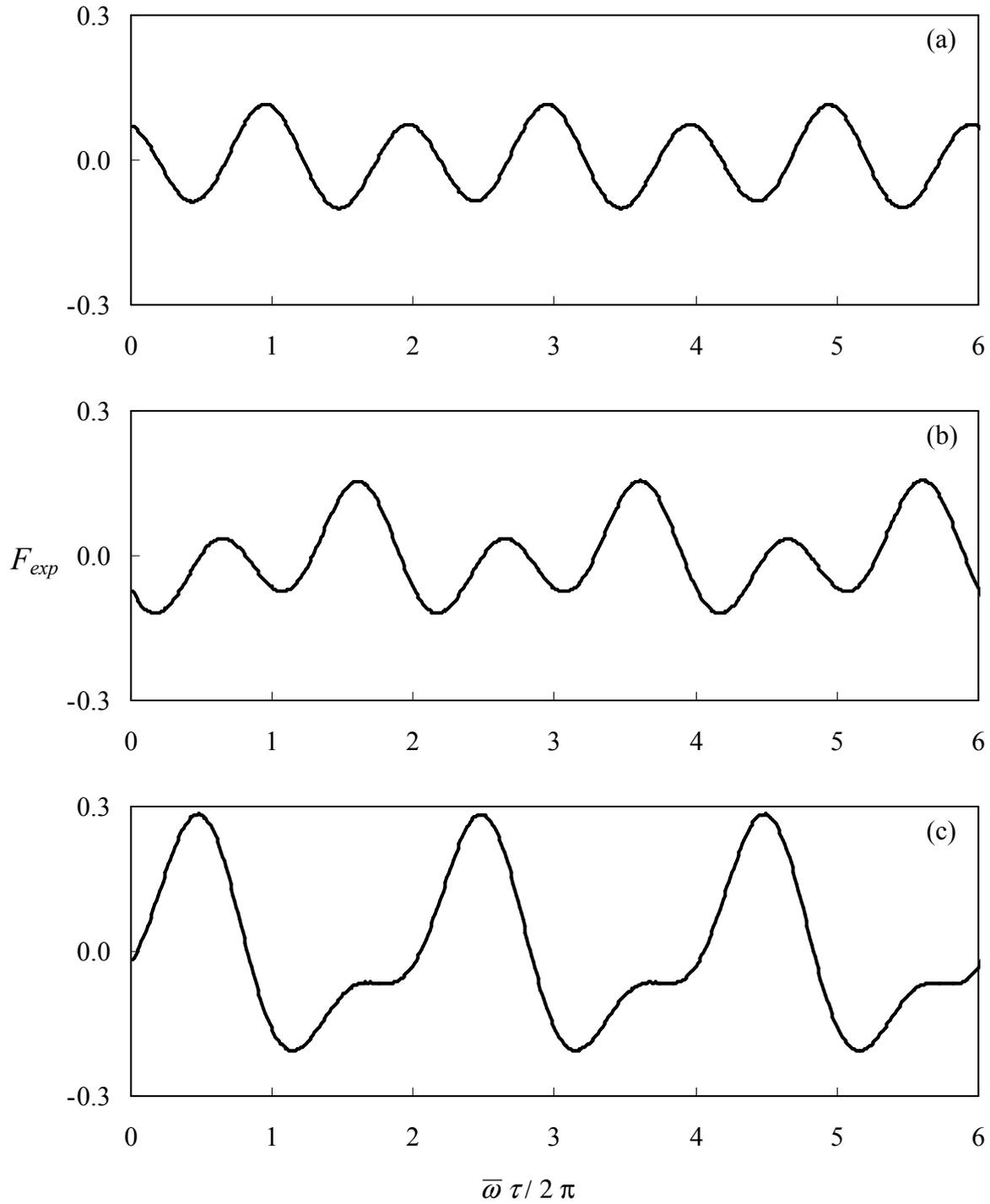

Figure 14. Time histories of the dimensionless normal force $F_{exp}$ versus the dimensionless time ($\overline{\omega} \tau / 2\pi$) for $\overline{\omega}$ = 2.023 (a), 2.020 (b) and 2.014 (c) and obtained for a dimensionless excitation force $\sigma \approx 30$ %.

J. PERRET-LIAUDET and E. RIGAUD.

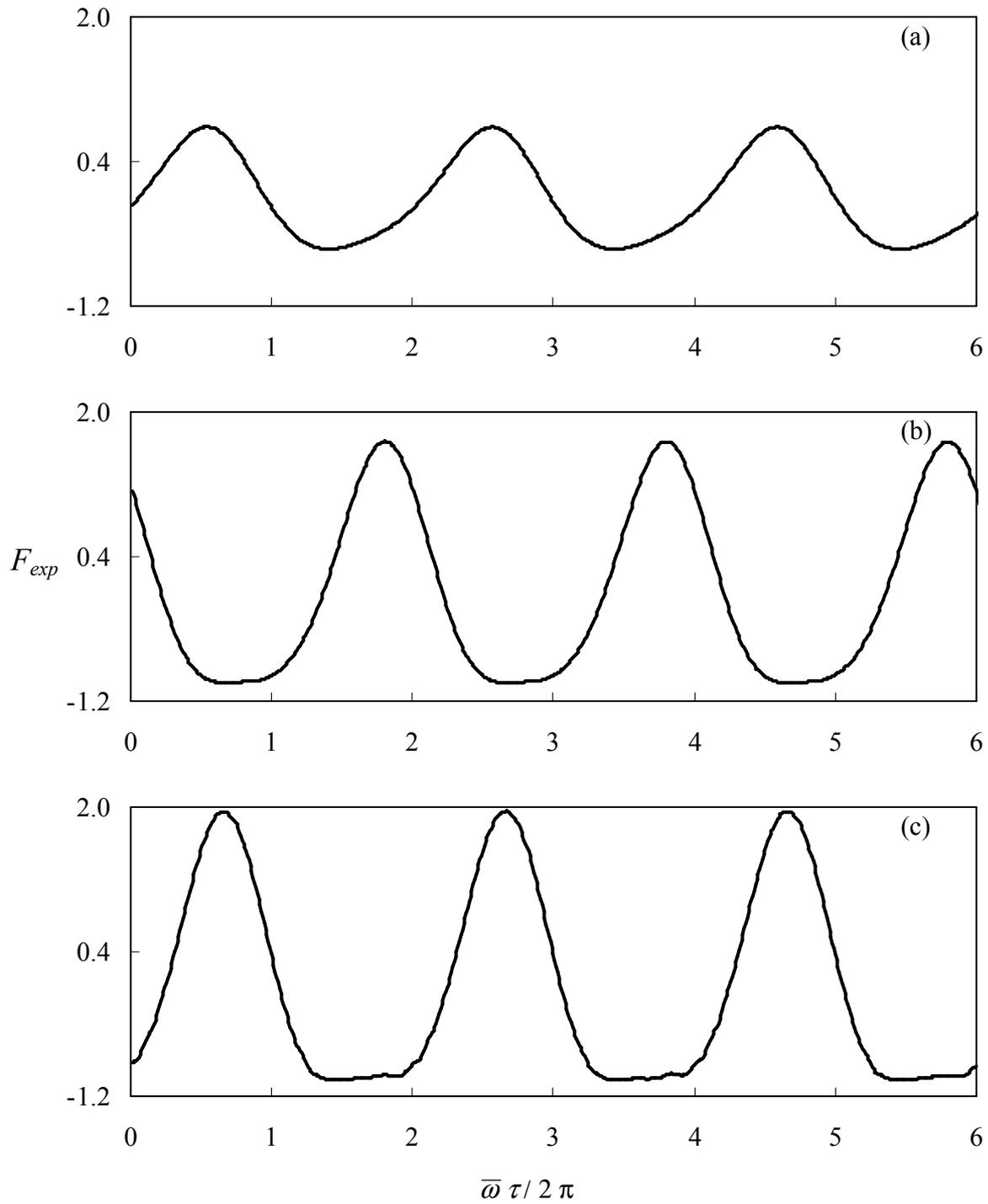

Figure 15. Time histories of the dimensionless normal force $F_{exp}$ versus the dimensionless time ($\overline{\omega}\,\tau/2\pi$) for $\overline{\omega}$ = 1.996 (a), 1.914 (b) and 1.848 (c) obtained for a dimensionless excitation level force $\sigma \approx 30\,\%$.

J. PERRET-LIAUDET and E. RIGAUD.